\documentclass[onecolumn,showpacs,amsmath,amssymb,11pt]{revtex4}
\usepackage{amsmath,amssymb,amsthm,bm,dcolumn,pifont,verbatim}

\def\at{\hat{a}_{\ell}^{\dag}}\def\a{\hat{a}_{\ell}}
\newcommand{\aat}[1]{\ensuremath{\hat{a}_{#1}^{\dag}}}
\newcommand{\ket}[1]{\ensuremath{\left|#1\right\rangle}}
\newcommand{\bra}[1]{\ensuremath{\left\langle #1\right|}}
\newcommand{\bracket}[2]{\ensuremath{\left\langle #1|#2\right\rangle}}

\newcommand{\deriva}[3]{\ensuremath{\frac{d^{#1}#2}{d #3^{#1}}}}
\def\be{\begin{equation}}\def\ee{\end{equation}}\def\vl{v_{\ell}}\def\l{\ell}\def\H{\Hat{H}_{\ell}}\def\A{\Hat{\mathfrak{A}}_{\ell}}\def\B{\Hat{\mathfrak{B}}_{\ell}}\def\HH{\Hat{H}}\def\AA{\Hat{\mathfrak{A}}}\def\BB{\Hat{\mathfrak{B}}}\def\o{\ket{^{\circ}}}\def\p{\bm{p}}\def\h{\hbar}\def\sech{\text{sech}}\def\x{\hat{x}}\def\p{\hat{p}}\def\uno{\hat{1}}\def\bes{\begin{subequations}}\def\ees{\end{subequations}}
\begin{document}
\title{Levinson's theorem and reflectionless one-dimensional potentials}
\author{D. E. Zambrano}
\email{dezambranor@unal.edu.co}
\affiliation{Departamento de F\'\i sica\\Universidad Nacional de Colombia, Sede Bogot\'a}
\date{October 3, 2005}    
\begin{abstract}
We studied two possible approaches to one-dimensional Levinson's theorem for Sch\"odinger equation, by using the reflectionless potential problem. In the first one, we restrict the 3-dimensional theorem in a direct way. In the other one, we use the theorem proposed by Dong, Ma and Klauss \cite{Dong}. We find failures in both of two approaches. In order to see this, we explicitly evaluate the phase shift using Schr\"odinger equation. A solution of equation is obtain by means a procedure proposed by Jaffe \cite{Jaffe}. \newline\newline
\end{abstract}
\maketitle
\section{Introduction}\label{Intro}
If we consider a potential with the form\cite{Jaffe}:
\be\label{vl}
\vl=-\l(\l+1)\sech^2x,
\ee
where $\l\geq0$ is an integer, transmission probability across this potential is 1. These kind of potential are known as \textit{reflectionless potential}. Schr\"odinger's equation with this kind of potentials has an exact solution (cf. \cite{Jaffe}).
In this paper we compare the result of Levinson's theorem with the solution of Schr\"odinger equation. In section \ref{acotados} we find the solution of this system and the phase shift for semi-bounded states, as in \cite{Jaffe}. In section \ref{levinson} restrict the 3-dimensional theorem and apply its result to the system. We find that yields a wrong result. In section \ref{Paridad} we test the theorem  proposed by  Ma, Dong and Klauss\cite{Dong}, finding the same inconsistence. Finally we discuss the results in section \ref{conc}.
\section{Phase shift for zero momentum}\label{acotados}
Following the method proposed by Jaffe \cite{Jaffe}, we find the phase shift for bounded state with zero momentum, so-called \emph{semi-bounded} states.
Beginning from Schr\"odinger equation
\be
\left[-\frac{\h^2}{2m}\deriva{2}{}{r}-V_0\sech^2(br)\right]\Psi(r)=E\Psi(r),
\ee
we write this equation for dimensionless parameters by means
\be
x=br,\quad v_0=\frac{2mV_0b^2}{\h^2},\quad k^2=\frac{2mb^2E}{\h^2}.
\ee
taking $v_0=\l(\l+1)$ where $l\geq0$ is an integer, we obtain a potential shape  as in Eq. \eqref{vl},
\be\label{H}
\H\Psi(x)=\left[\p^2-\l(\l+1)\sech^2x\right]\Psi=k^2\Psi(x).
\ee
So we have
\begin{itemize}
\item For $k^2>0$ a scattering state,
\item For $k^2\geq0$ and $||\Psi||<\infty$ a bounded state,
\end{itemize}
where $\Psi$ is the wave-function.
In accordance with \cite{Jaffe}, we define the operators
\begin{align}
\a&=\p-i\l\tanh x,& \at&=\p+i\l\tanh x.
\end{align}

Since $[\x,\p]=i$, is clear that
\begin{align}\label{A y B}
\A&\equiv\at\a=\HH_\l+\uno\l^2,&
\B&\equiv\a\at=\HH_{\l-1}+\uno\l^2.
\end{align}
Now, we define the state $\ket0_{\l}$ by $\a\ket{0}_{\l}=\o,$ where $\o$ is \emph{null} ket. Denoting $\Psi^{(0)}_{\l}(x)=\bracket{x}{0}_{\l}$, we obtain
\be
\left[-i\deriva{}{}{x}-i\l\tanh x\right]\Psi_{\l}^{(0)}(x)=0.
\ee 
The solution for this equation is 
\begin{equation}\label{psi ground}
\Psi_{\l}^{(0)}(x)=N_{\l}\,\sech^{\l}(x),\qquad\l\geq0,
\end{equation}
 where $N_{\l}$ is a normalisation constant. Due to linearity of the operator $\at$,
$
\A\ket0_{\l}=0\ket0_{\l}, $
then $\ket0_{\l}$ is an eigenstate of $\A$ associated to the zero eigenvalue, and is the ground state because does not have any node. In the other hand, if $\ket{\Psi}$ is an eigenstate of  $\A$ associated to the eigenvalue $\psi\neq0$ we have 
\be
\a\Bigl(\A\ket{\Psi}\Bigr)=\left(\a\at\right)\a\ket{\Psi}=\B\Bigl(\a\ket{\Psi}\Bigr)=\psi\Bigl(\a\ket{\Psi}\Bigr),
\ee
then if $\psi\neq0$, $\ket{\Psi'}=\a\ket{\Psi}$ is an eigenvalue of $\B$ associated to the \textit{same eigenvalue} $\psi$. Explicitly
\bes
\begin{align}
\A\ket{\Psi}&=\H\ket{\Psi}+\l^2\ket{\Psi}=\psi\ket{\Psi},\\
\B\ket{\Psi'}&=\HH_{\l-1}\ket{\Psi'}+\l^2\ket{\Psi'}=\psi\ket{\Psi'},\quad\psi\neq0.
\end{align}
\ees
Therefore
\bes\label{Eq vp Hl y Hi-1}
\begin{align}
\H\ket{\Psi}&=\Bigl(\psi-\l^2\Bigr)\ket{\Psi}\label{Eq vp Hl},\\
\HH_{\l-1}\ket{\Psi'}&=\Bigl(\psi-\l^2\Bigr)\ket{\Psi'},\quad\psi\neq0\label{Eq vp Hl-1}
\end{align}
\ees
thus $\ket{\Psi}$ and $\at\ket{\Psi}$ form the set of eigenstates of $\H$ y $\HH_{\l-1}$, respectively, and these Hamiltonians have the same spectrum, except  by the eigenvalue associated to the ground state of $\H$. This is a very important result obtained by Jaffe \cite{Jaffe}.
\newline\newline
For $\l=0$, we have that $\HH_0=\p^2$, i.e. free particle case, a well-known system. Energy eigenstates are
\be
\Psi_0(k,x)\equiv\bracket{x}{k}_0=\exp(ikx),
\ee
where subindex zero indicates $\l=0$. Energy eigenvalues are $E(k)=k^2$, since we are woking with dimensionless quantities.

For $\l=1$, we construct the eigenfunctions of $\HH_1$ for a given $k^2$.
We define \mbox{$\ket{k}_1=\aat1\ket{k}_0$.} If we apply $\HH_1$ to $\ket{k}_1$ we obtain
\be\label{estasi}
\HH_1\ket{k}_1=(\AA_1-\uno)\aat1\ket{k}_0=\aat1\BB_1\ket{k}_0-\aat1\ket{k}_0
=(k^2+1)\aat1\ket{k}_0-\aat1\ket{k}_0=k^2\aat1\ket{k}_0=k^2\ket{k}_1.
\ee
therefore $\ket{k}_1$ is an eigenstate of $\HH_1$ associated to the eigenvalue $k^2$, as we hope, and so the wave function is
\be\label{psi1}
\Psi_1(k,x)=\bracket{x}k_1=\bra{x}\aat1\ket{k}_0
=(k+i\tanh x)\exp(ikx).
\ee
In the region \textit{before}  to potential well, i.e. for $x\to-\infty$, we may express the wave-function as an asymptotic incident function $I_1(k)e^{ikx}$ and a reflected wave function $R_1(k)e^{-ikx}$. In the same manner, in the region \textit{after} to the potential well, i.e. $x\to\infty$, we may write the wave-function as $T_1(k)e^{ikx}$, the transmitted part, 
\be\label{def R y T}
\lim_{x\to-\infty}\Psi_{1}(k,x)=I(k)e^{ikx}+R_{1}(k)e^{-ikx}
,\quad\quad
\lim_{x\to\infty}\Psi_{1}(k,x)=T_{1}(k)e^{ikx}
\ee
From Eq. (\ref{psi1}) we can see that
\begin{align}
R_1(k)&=0,&
T_1(k)&=k+i,&
I_1(k)&=k-i.
\end{align}
At this moment we can observe several aspects. First, since the reflection coefficient $R(k)$ is null, there not exist reflected wave function, i.e. reflectionless. Second, the ratio between incident and transmitted waves is
\be
\frac{T(k)}{I(k)}=\frac{k+i}{k-i}=\exp\left(2i\arctan\left(\frac1k\right)\right).
\ee
This indicates that the interaction with the potential only yields a phase shift, therefore the transmition probability is 1.
Phase shift is defined as \cite{Nw66}
\be
\delta_{\l}=\frac12\arg\left(\frac{T(k)}{I(k)}\right),
\ee
so, for $\ell = 1$ we have,
\be
\delta_1=\arctan\left(\frac1k\right).
\ee
Repeating the same procedure $\l$ times, eigenfunctions can be constructed by means
\be\label{psi l}
\Psi_{\l}(k,x)=\aat{\l}\aat{\l-1}\dots\aat1\Psi_0(k,x)
=\left[\prod_{j=1}^{\l}(k+ij\tanh x)\right]\exp(ikx),
\ee 
Last expression allows us to obtain
\begin{align}
R_{\l}(k)&=0,&
T_{\l}(k)&=\prod_{j=1}^{\l}(k+ij),&
I_{\l}(k)&=\prod_{j=1}^{\l}(k-ij).
\end{align}

Hence, for any $\l$ there not exist reflected wave and 
\be
\frac{T_{\l}(k)}{I_{\l}(k)}=\cfrac{\prod\limits_{j=1}^{\l}(k+ij)}{\prod\limits_{j=1}^{\l}(k-ij)}=\exp\left(2i\sum_{j=1}^{\l}\arctan\left(\frac{j}{k}\right)\right)
\ee
indicating only a phase shift of $\delta_{\l}(k)$ given by
\be
\delta_{\l}(k)=\sum_{j=1}^{\l}\arctan\left(\frac{j}{k}\right)
\ee
Taking the limit $k\to0$, i.e. for zero momentum, we have
\be\label{shift}
\delta_{\l}(0)=\lim_{k\to0}\sum_{j=1}^{\l}\arctan\left(\frac{j}{k}\right)\\
=\frac{\pi}2\l
\ee
\section{Levinson's theorem}\label{levinson}
Levinson's theorem \cite{Nw66} establishes that the phase shift $\delta_{\l}(0)$ for states of zero momentum y and angular momentum $\l$ for 3-dimensional Schr\"odinger equation with a spherical symmetric potential is given by
\be\label{Lev3d}
\begin{split}\
\delta_{\l}(0)=\begin{cases}
n_{\l}\pi+\pi/2&\text{critical case}\\
n_{\l}\pi&\text{no critical case}
\end{cases}
\end{split}
\ee
where $n_{\l}$ is the number of bounded states with angular momentum $\l$. Critical case means existence of semi-bounded state (also energy zero or half-bound state), i.e. $k^2=0$, but wave function is finite and not necessarily square integrable. Potential $V(r)$ must fulfills the asymptotic  condition:
\be
r^2|V(r)|\to0,\quad r\to\infty.
\ee
We would restrict this theorem to one-dimensional case, supposing an even potential (symmetric in one dimension) which fulfills:
\be\label{CondiLev}
x^2|V(x)|\to0,\quad x\to\infty.
\ee
For the potential (1), the conditions of restricted theorem are satisfied, since 
\be
\lim_{x\to\infty}x^2v_{\l}(x^2)=-4\l(\l+1)\lim_{x\to\infty}\frac{x^2}{\left(e^x+e^{-x}\right)^2}=0
\ee
In conclusion the restricted theorem is applicable to reflectionless potentials. For $\l=0$, in addition to the plane waves, i.e. the eigenfunctions for free particle, it must exist a the ground state $\ket0_0$ determined by
\be
\deriva{}{\Psi_0^{(0)}(0,x)}x=0
\ee
hence $\Psi_0(0,x)=N_0$, is a constant and is associated to eigenvalues $E_0^{(0)}=0$, then it is a semi-bounded state and since there not exist bounded states\cite{Dong}, this state represents the critical case for \eqref{Lev3d}, then the restricted theorem states
\be
\delta_0=\frac{\pi}2.
\ee
This result contradicts \eqref{shift}. Therefore is not valid restrict the theorem \eqref{Lev3d} to the one-dimensional case.
\section{Levinson's theorem for wave function with parity}\label{Paridad}
In the last section we show that restrict the 3-dimensional theorem is not valid. In 3-dimensional case we have an additional ingredient: the angular momentum, which is lost if we restrict to a one-dimensional case in a direct way.
Ma, Dong and Klauss\cite{Dong} found a  restriction, in principle valid, for the Levinson theorem. This restriction affirms for the one-dimensional case:
 
For one-dimensional Schr\"odinger equation with an even potential $V(x)$ which fulfills \eqref{CondiLev} the phase shift for a state with definite parity at zero momentum is
\be\label{LEV}
\begin{aligned}
\left.
\begin{array}{cl}
\delta_{e}(0)&=n_{e}\pi+\pi/2,\\
\delta_{o}(0)&=n_{o}\pi
\end{array}\right\}&\text{non-critical case,}\\
\left.
\begin{array}{cl}
\delta_{e}(0)&=n_{e}\pi\\
\delta_{o}(0)&=n_{o}\pi+\pi/2,
\end{array}\right\}&\text{ critical case,}
\end{aligned}
\end{equation}
where $n_e$ and $n_o$ are the number of bounded states for the even and odd wave-function, respectively and $\delta_e(0)$ and $\delta_o(0)$ are the phase shift at zero momentum for even and odd case. Critical case occurs if there exist a semi-bounded state.
\newline
\newline For $\l=0$, the wave function at zero momentum is $\Psi_0^{(0)}(x)$, which is an even function. In the other hand, we have a semi-bounded state and not bounded ones, then \eqref{LEV} affirms $\delta_0=0$ which is true.\newline
Now, we analyse the $\l=1$ case. The spectrum of the Hamiltonian is the same than $\l=0$ except for an eigenvalue associated to the state $\ket0_1$, which is evaluated using \eqref{Eq vp Hl} for $\psi=0$,
\be
\HH_1\ket{0}_1=-\ket{0}_1.
\ee
then $E_1^{(0)}=-1$, a bounded state. In order to obtain the wave function $\Psi_1^{(0)}(x)$ we replace $\l=1$ in Eq. \eqref{psi ground},
\be
\Psi_1^{(0)}(x)=\bracket{x}0_1=N_1\sech x.
\ee
This function is even. Replacing $\Psi_0^{(0)}$ in \eqref{psi l} we obtain the zero energy (semi-bounded) wave function 
\be
\Psi_1^0(x)=iN_0\tanh x
\ee
whose parity is odd. There are not any odd bounded states, thus the phase shift determined by \eqref{LEV} is
 $\delta_1(0)=\pi/2,$ yielding the same phase shift than \eqref{shift} for $\l=1$. Note that the bounded state $\ket0_1$ does not contribute to the phase shift, since have a distinct parity than the half-bounded state.

In the case $\l=2$, we have a bounded state for $E=-1$, and its corresponding eigenfunction is
\be
\Psi_2^{(1)}=\bra{x}\hat{a}^{\dag}_2\ket{0}_1\propto (k+2i\tanh x)\sech x\propto\frac{\tanh x}{\cosh x},
\ee
which is an odd function. Additionally, we have a bounded state for $E_2^{(0)}=-4$ given by
\be
\Psi_2^{(0)}=N_2\sech^2x
\ee
which is even. Also we obtain an even semi-bounded state for $k^2=0$ given by
\be
\Psi_2^{0}(x)=-N_0\tanh^2 x
\ee
Hence, there are two bounded states, one odd and other even, and an even semi-bounded state associated to $k^2=-1$, then \eqref{LEV} establishes that $\delta_\l(0)$ is a semi-integer times $\pi$, contradicting the phase shift given by the Sch\"odinger equation (see Eq. \eqref{shift}), which value is $\pi$. 
\section{Discussion}\label{conc}
Although the potential \eqref{vl} fullfils the conditions of the \emph{restricted} theorem, in his two versions, the direct restriction fails from $\l=0$, instead the approach of Ma, Dong and Klauss fails from $\l=2$.

Is clear that Schr\"odinger equation provides the \emph{actual} result for phase shift, it must exist a \emph{hide} or ignored  condition in the demonstration, in the case of \cite{Dong}. In addition, is clear that in one-dimensional case the angular momentum is not considered, since have not one-dimensional sense . The idea proposed in \cite{Dong} is well method to include the lack of angular behaviour, but is not enough, some class of imformation is lost in the process.
%
%
%
%

\end{document}